\documentclass[aps,pra,twocolumn,notitlepage]{revtex4-1}
\usepackage{graphicx,amsfonts,amssymb,amsmath,hyperref,hypcap,enumerate,enumitem}
\usepackage{xcolor}
\usepackage{soul}
\usepackage{scalerel}
\usepackage{adjustbox,stmaryrd}
\usepackage{cancel}
\usepackage{csquotes}
\usepackage[utf8]{inputenc}
\usepackage{geometry}
\hypersetup{  colorlinks=true, linkcolor=blue, citecolor=blue, urlcolor=blue  }

\begin{document}
\title{2D bilayer electron-hole superfluidity with unequal and anisotropic masses}
\author{Jihang Zhu}
\affiliation{Condensed Matter Theory Center and Joint Quantum Institute, Department of Physics, University of Maryland,
College Park, Maryland 20742, USA}
\author{Sankar Das Sarma}
\affiliation{Condensed Matter Theory Center and Joint Quantum Institute, Department of Physics, University of Maryland,
College Park, Maryland 20742, USA}

\begin{abstract}
We investigate the stability of electron–hole superfluidity in two-dimensional bilayers with unequal and anisotropic effective masses. Using a zero-temperature, self-consistent Hartree–Fock approach, we study two experimentally relevant deviations from the ideal equal-mass isotropic case: (i) isotropic but unequal conduction and valence band masses ($m_c^* \neq m_v^*$), and (ii) equal average masses with orthogonal in-plane anisotropies $(m_{c,x}^*, m^*_{c,y}) = (m_1^*, m_2^*)$ and $(m^*_{v,x}, m^*_{v,y}) = (m_2^*, m_1^*)$. For both scenarios, we compute the order parameter and analyze the BEC–BCS crossover as a function of layer separation and mass ratio. We find that both mass imbalance and mass anisotropy reduce the pairing strength and suppress the inferred critical temperature $T_c$ by breaking perfect Fermi surface nesting, and shift the BEC-BCS crossover. Despite these effects, superfluidity remains robust across the full range of densities and interlayer separations considered, with no transition to an unpaired plasma state in the absence of screening. Our results provide a baseline for understanding the interplay of mass mismatch and anisotropy in current and emerging bilayer platforms, including van der Waals heterostructures and anisotropic two-dimensional semiconductors.
Our work also establishes that Fermi surface nesting is not a key ingredient for the bilayer superfluidity, which is always the ground state for all electron-hole bilayers although the resultant $T_c$ depends on the parameter details and may very well be unmeasurably low for large interlayer separations.
\end{abstract}

\maketitle

\section{Introduction}
The condensation of electron-hole pairs into a macroscopic quantum state, potentially exhibiting superfluidity, represents a fascinating manifestation of coherent many-body physics in solid-state systems. Two-dimensional (2D) bilayers, where electrons and holes are confined to spatially separated layers, have emerged as particularly promising platforms for realizing and studying this phenomenon \cite{LozovikYudson1975JETPL, LozovikYudson1976JETP, Shevchenko1976LowTemp, Shevchenko_PhaseDiagram_1994, LZheng_doubleQW_1997, LZheng_SDS_1997, SDS_LZheng_1997, SDS_LZheng_1998, Eugene_SDS_1999, Eugene_SDS_2000, FCWu_exciton_2015, DinhDuyVu_SDS_2023, JZhu_SDS_2024, JZhu_SDS_pseudospin_2024, Cookmeyer_SDS_2024, YTTu_SDS_2024, Sreejith_SDS_2024}. The Coulomb attraction between an electron in one layer and a hole in the other can lead to the formation of an interlayer exciton (a spatially indirect exciton), with the advantage of inhibiting tunneling between layers as a result of the insulating barrier in between, allowing such excitons to remain stable and--under suitable conditions--to condense into a superfluid state.
When interlayer tunneling is negligible \cite{Conti_ehSuperfluid_1998}, the resulting condensate can exhibit counterflow superfluidity--coherent, dissipationless motion of electron-hole pairs.

One route to excitonic condensation is through the excitonic insulator \cite{keldysh1965excitonic, jerome1967excitonic, Ma_EI_2021} phase, which can spontaneously form in systems with a negative band gap (semimetals with overlapping conduction and valence bands) or a small positive gap (semiconductors) that is smaller than the exciton binding energy. In this scenario, the normal semimetallic or semiconducting ground state becomes unstable towards the spontaneous formation of electron-hole pairs, opening a many-body gap (2$\Delta$) in the excitation spectrum.
Such bulk electron-hole systems are necessarily unstable since the electrons and holes occupy the same spatial region and eventually recombine or form electron-hole droplets, and the system is essentially not in equilibrium. In addition, such bulk electron-hole excitonic systems typically undergo lattice instabilities and charge density wave transitions instead of forming bosonic superfluid \cite{Kohn_1967, Kohn_BEC_1970, Halperin_1968}.

Most work on excitonic interaction-induced electron-hole superfluidity over the last 30 years has focused on bilayers where one layer has electrons and the other has holes, so the superfluidity is through interlayer excitonic coupling. The spatial separation of the electrons and holes prevents any recombination, and thus a true interlayer excitonic superfluid ground state should form at zero temperature. An equivalent system with similar, but not identical physics is a quantum Hall bilayer at the complete filling of the lowest Landau level (so that each layer is half-filled by electrons), where the electron-hole symmetry of quantum Hall physics ensures that the system is equivalent to an electron-hole bilayer \cite{sarma2008perspectives, sarma2008perspectives_Girvin, sarma2008perspectives_Eisenstein, Eisenstein_ExcitonCond_2004, Eisenstein_exciton_2014, Eisenstein_2DEGinB_2004, Tutuc_GaAs_exciton_2004, Eisenstein_science_2004}.
Our focus in the current work is on the zero-field bilayer situation where the two layers are occupied by electrons and holes, providing a classic electron-hole interlayer superfluidity scenario without any finite magnetic field quantum Hall physics complications.

The earliest experimental platforms of interlayer exciton physics and the search for superfluidity were GaAs/AlGaAs coupled quantum wells (QWs) \cite{LZheng_doubleQW_1997, LZheng_SDS_1997, Fukuzawa_CQWs_1990, Kash_CQWs_1991, Sivan_CQWs_1992, Kane1994APL, Butov_CQWs_1994, XZhu_QW_1995, Gupta_QW_2011, SDS_LZheng_1997, SDS_LZheng_1998, Eugene_SDS_1999, Eugene_SDS_2000}, where independent electrostatic control of electron and hole densities was possible. These systems were used to probe excitonic correlations through Coulomb drag \cite{Croxall_CoulombDrag_2008, Seamons_CoulombDrag_2009, Tiemann_ExcitonCond_2008, Nandi_CoulombDrag_2012, Eisenstein_ExcitonCond_2004, Nguyen_CoulombDrag_2025, Gamucci_2014, JIALi_coulombDrag_2016, Tutuc_drag_2016, XLiu_ExcitonSuperfluid_2022, XLiu_drag_2017, JIALi_ExcitonSuperfluid_2017, Ma_EI_2021, Qi_eh_2023, Davis_coulombDrag_2023, Kim_coulombdrag_2001, Hwang_SDS_2008, Ho_CoulombDrag_2018} and counterflow \cite{JJSu_2008, Joglekar_2005} transport, motivated by predictions that dissipationless counterflow currents would serve as a clear signature of superfluidity. While such measurements provided evidence for strong interlayer correlations, the relatively low exciton binding energies and disorder effects limited critical temperatures, preventing unambiguous observation of superfluidity.
One question remaining open is, although there is certainly a finite-temperature phase transition at a critical temperature $T_c$ driving the system from superfluid to normal, whether there is also a $T=0$ quantum phase transition as a function of layer separation driving a superfluid phase at small layer separations to a normal phase  (i.e. a U(1) breaking quantum phase transition) at larger separations.  In the current work, we show that there is no such $T=0$ critical transition at a critical separation, and the system is always a superfluid at $T=0$ albeit with exponentially low temperatures for larger separations, making the observation of the superfluid phase a challenge.

The emergence of 2D materials such as graphene, transition metal dichalcogenides (TMDs), and phosphorene has provided several advantages over traditional semiconductor QWs.
Reduced dimensionality enhances Coulomb interactions by weakening dielectric screening, producing exciton binding energies that can be hundreds of meV in TMDs. Additional internal degrees of freedom, such as valley pseudospin in graphene and TMDs, and intrinsic anisotropy in materials like phosphorene open new directions for engineering excitonic condensates. In van der Waals heterostructures, such as double (bi)layer graphene \cite{XMou_doubleG_2015} or MoSe$_2$/WSe$_2$ bilayers \cite{ZWang_highT_2019, Nguyen_CoulombDrag_2025}, the combination of tunable band gaps, large masses, and moir\'e superlattices has pushed predicted transition temperatures to the liquid-nitrogen range or beyond.
Often, these new 2D materials have much lower background lattice dielectric constants than the GaAs-based bilayers, making the effective interlayer Coulomb coupling much stronger. Additionally, these novel van der Waals bilayers typically can be fabricated with much smaller interlayer separations ($\sim 1$ nm) providing strong interlayer excitonic effects. Recent bilayer excitonic superfluity research has focused more on these van der Waals based bilayers although work still continues in the GaAs bilayers too.

Excitonic condensation in bilayers can be viewed as a solid-state analogue of Bose-Einstein condensation (BEC), enriched by Fermi statistics, long-range Coulomb forces, and reduced dimensionality. 
In the low-density limit, tightly bound excitons behave as dilute composite bosons undergoing BEC \cite{LZhang_BEC_2024}; in the high-density limit, the system resembles a BCS-like state of weakly bound, overlapping electron-hole pairs.
Importantly, true superfluidity manifests through dissipationless counterflow currents, equal in magnitude and opposite in direction in the two layers--a hallmark distinct from ordinary charge transport.
Screening effects, especially strong in the high-density BCS regime, tend to suppress pairing, while in the low-density BEC regime mean-field theory may overestimate stability.
It is, however, important to emphasize that the bilayer system remains an interlayer superfluid condensate with the spontaneous broken U(1) layer symmetry throughout the BEC-BCS crossover from the strongly-coupled BEC regime at smaller separations with high $T_c$ to the weakly-coupled BCS situation with lower $T_c$ as the layer separation increases.  There is no zero-temperature phase transition at any layer separation.

Early and recent attention has turned to how deviations from the ideal equal-mass isotropic \cite{DePalo_symmetric_2002} case affect superfluidity \cite{Pieri_BCSBEC_2007, Kaneko_ED_2013}. Mass imbalance \cite{Kaneko_ED_2013}, present for example in GaAs where $m_h/m_e \approx 6–7$, breaks perfect Fermi-surface nesting and suppresses pairing in the BCS limit. Anisotropic effective masses, as in phosphorene or certain transition metal dichalcogenides, produce direction-dependent superfluid stiffness, coherent length, and Kosterlitz-Thouless transition temperatures. Prior studies \cite{Zittartz_eh_MassAniso_1967, Conti_ehSuperfluid_1998} have shown that anisotropic electron or hole band is deleterious to superfluidity, and in some mean-field treatments introduce a minimum threshold coupling for superfluidity \cite{Conti_ehSuperfluid_1998}, in contrast to the standard BCS picture.

In this work, we address two general and experimentally relevant scenarios that disrupt perfect Fermi-surface nesting in 2D electron-hole bilayers with equal electron and hole densities. The first is mass imbalance, where electrons and holes have isotropic but unequal effective masses ($m_e^* \neq m_h^*$), introducing an energy mismatch between the nested Fermi surfaces. The second is anisotropy with equal average masses, where electrons and holes have orthogonal anisotropies $(m^*_{e,x}, m^*_{e,y}) = (m_1^*, m_2^*)$ and $(m^*_{h,x}, m^*_{h,y}) = (m_2^*, m_1^*)$, destroying nesting through geometric distortion of the Fermi surfaces. Using a zero-temperature self-consistent Hartree-Fock (HF) framework without screening, we map the phase diagram across a wide range of mass ratios and anisotropy parameters. We show how both mass imbalance and anisotropy influence the order parameter and the strong-to-weak coupling BEC-BCS crossover \cite{Kaneko_ED_2013}. Within the regime studied, we find that superfluidity persists for all parameters considered--there is no transition to an unpaired electron-hole plasma as the interaction is reduced by increasing layer separation--though the pairing strength and inferred $T_c$ are strongly suppressed.
Our detailed theoretical consideration of the role of anisotropy in the bilayer excitonic condensation is of qualitative, and not just quantitative, significance because it directly addresses the extent to which the Fermi surface mismatch plays a qualitative role in the bilayer supefluidity phenomenon.  For example, our finding that the non-existence of any zero temperature quantum phase transition (and only a smooth crossover) as a function of layer separation survives the absence of Fermi surface mismatch points to the fundamental superfluid ground state in the bilayer electron-hole system at $T=0$ independent of any Fermi surface nesting or not.

For convenience, we will use the language of conduction band (for electrons) and valence band (for holes) in the rest of the paper, their corresponding effective masses will be labeled by $m^*_c$ for electrons and $m^*_v$ for holes.
In Sec.~II, we develop the basic self-consistent HF theory applying the formalism to produce results for unequal electron and hole masses in the isotropic situation.  In Sec.~III we generalize the theory to geometrically non-nested Fermi surfaces with anisotropic masses.  We conclude with a discussion in Sec.~IV.  Detailed numerical results are presented in subfigures as 8 panels each in six figures.

\begin{figure*}[!htb]
\centering
\includegraphics[width=1.0\textwidth]{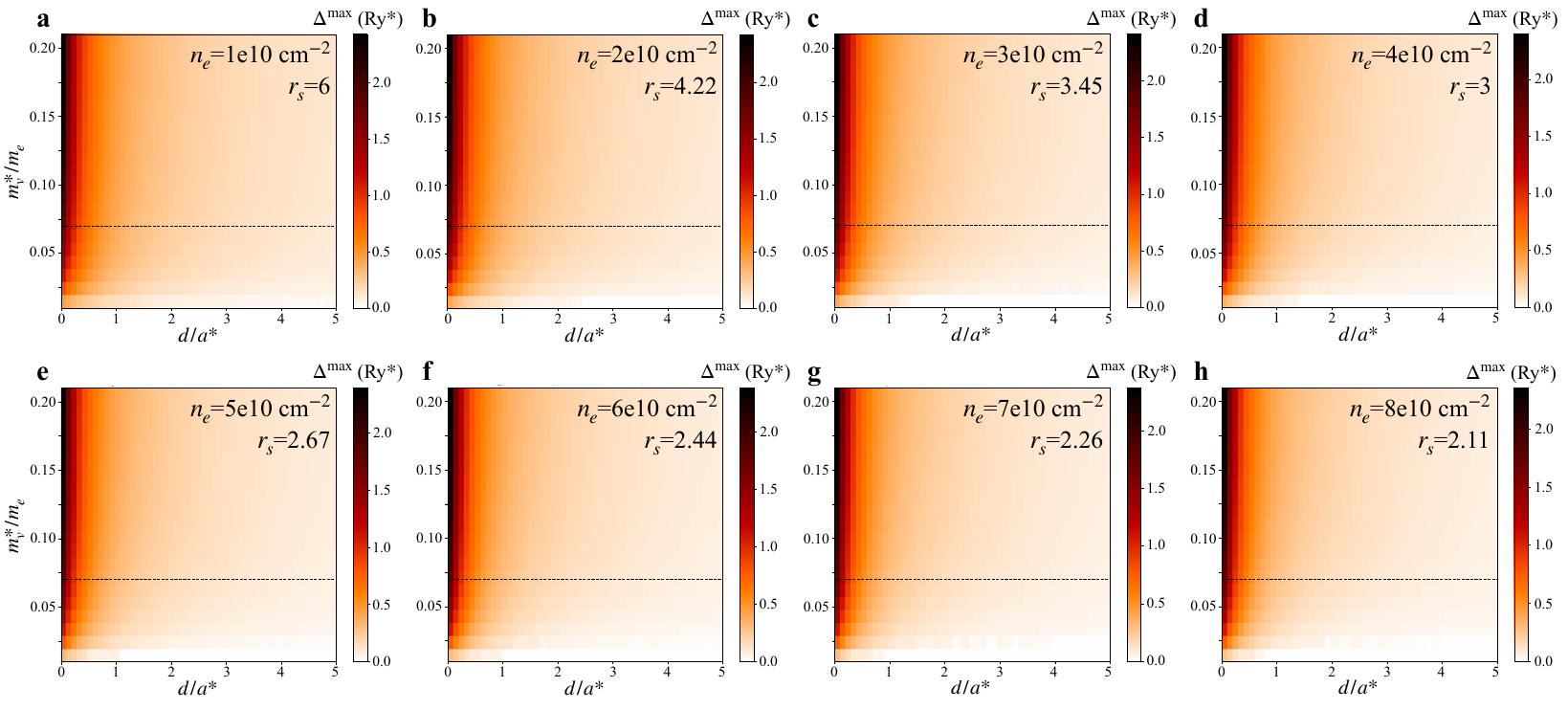}
\caption{\label{fig1_diffmv_2d} { {\bf{Self-consistent HF at $T=0$ for unequal and isotropic masses, $m_c^* \neq m_v^*$, with $m_c^* = 0.07 m_e$ fixed.}}
Color maps show the maximum order parameter $\Delta^{\rm max} \equiv \max\{ \Delta_{\mathbf{k}}\}$ (in unit of Ry$^*$) as a function of interlayer spacing $d/a^*$ and valence band mass $m_v^*/m_e$ for electron densities $n_e \in [1, 8] \times 10^{10}$ cm$^{-2}$.
The black dashed line in each figure marks the equal-mass case $m_v^* = m_c^* = 0.07m_e$.
Here, Ry$^*$ and $a^*$ are defined using $m^* = 0.07 m_e$.
}}
\end{figure*}

\section{HF with unequal and isotropic masses}
We consider an electron-hole bilayer where electrons (holes) come from a single parabolic conduction (valence) band. The spinless HF Hamiltonian \cite{Comte_ehHF_1982} is given by
\begin{equation}
\label{Eq_HamilHF}
\begin{split}
\hat{\mathcal{H}}_{\rm HF} &= \sum\limits_{\mathbf{k}}
\begin{pmatrix}
c^\dagger_{c \mathbf{k}} & c^\dagger_{v \mathbf{k}}
\end{pmatrix} 
\begin{pmatrix}
\varepsilon_{c \mathbf{k}} & -\Delta_\mathbf{k} \\
-\Delta^*_\mathbf{k} & \varepsilon_{v \mathbf{k}}
\end{pmatrix}
\begin{pmatrix}
c_{c \mathbf{k}} \\
c_{v \mathbf{k}}
\end{pmatrix}.
\end{split}
\end{equation}
The off-diagonal term, $-\Delta_{\mathbf{k}}$, represents the electron-hole attraction induced by interlayer exchange interaction.
The diagonal term, $\varepsilon_{c \mathbf{k}}$ ($\varepsilon_{v \mathbf{k}}$), is the conduction (valence) band dispersion normalized by the Hartree energy $V_H$ and the intralayer exchange energy $V_{x}(\mathbf{k})$:
\begin{equation}
\label{Eq_epcv}
\begin{split}
\varepsilon_{c \mathbf{k}} &= \varepsilon^{(0)}_{c \mathbf{k}} + V_{H,c} + V_{x,c}(\mathbf{k}), \\
\varepsilon_{v \mathbf{k}} &= \varepsilon^{(0)}_{v \mathbf{k}} + V_{H,v} + V_{x,v}(\mathbf{k}).
\end{split}
\end{equation}
$\varepsilon_{c\mathbf{k}}^{(0)}$ and $\varepsilon_{v\mathbf{k}}^{(0)}$ are bare parabolic dispersions with effective masses $m_c^*$ and $m_v^*$,
\begin{equation}
\label{Eq_varepsilon0}
\varepsilon^{(0)}_{c \mathbf{k}} = \frac{\hbar^2 k^2}{2m_c^*},
\quad 
\varepsilon^{(0)}_{v \mathbf{k}} = -\frac{\hbar^2 k^2}{2\tilde{m}_v^*} -E_g.
\end{equation}
Here, $E_g$ denotes the conduction-valence band overlap, which is determined by the initial electron density $n_e$ in our calculation.
In Eq.~(\ref{Eq_varepsilon0}), $\tilde{m}_v^*$ is the renormalized valence band effective mass which accounts for exchange interactions from all occupied states
\begin{equation}
\label{Eq_tilde_mv}
\begin{split}
-\frac{\hbar^2 k^2}{2\tilde{m}_v^*} 
&= -\frac{\hbar^2 k^2}{2m_v^*} - \frac{1}{A} \sum\limits_{\mathbf{k}'} V^{S}_{\mathbf{k}-\mathbf{k}'} \langle c^\dagger_{v\mathbf{k}'} c_{v\mathbf{k}'} \rangle_0 \\
&= -\frac{\hbar^2 k^2}{2m_v^*} - \frac{1}{A} \sum\limits_{\mathbf{k}'} V^{S}_{\mathbf{k}-\mathbf{k}'} \rho_{vv}^0(\mathbf{k}'),
\end{split}
\end{equation}
where $\rho_{vv}^0(\mathbf{k}') = \langle c^\dagger_{v\mathbf{k}'} c_{v\mathbf{k}'} \rangle_0 = 1$ is the expectation value in the reference state where all valence band states are occupied and all conduction band states are empty \cite{YPShim_exciton_2009}.
The density matrix element $\rho_{vv}(\mathbf{k})$ is defined relative to this reference state expectation, and we denote it as
\begin{equation}
\tilde{\rho}_{vv}(\mathbf{k}) = \rho_{vv}(\mathbf{k}) - \rho^0_{vv}(\mathbf{k}).
\end{equation}
The Hartree terms in Eq.~(\ref{Eq_epcv}) are given by
\begin{equation}
\begin{split}
V_{H,c} &= -V_{H,v} \\
&= \frac{\pi e^2 d}{A\epsilon} \sum\limits_{\mathbf{k}'} \Big( \rho_{cc}(\mathbf{k}') - \tilde{\rho}_{vv}(\mathbf{k}') \Big). \\
\end{split}
\end{equation}
For equal electron and hole densities that we consider in this paper,
\begin{equation}
\label{Eq_VH}
\begin{split}
V_{H,c} &= -V_{H,v} = \frac{2\pi e^2 d n_e}{\epsilon}, 
\\
\end{split}
\end{equation}
where
\begin{equation}
n_e = \frac{1}{A} \sum\limits_{\mathbf{k}} \rho_{cc}(\mathbf{k})
\end{equation}
is the electron density.
The intralayer exchange energies in Eq.~(\ref{Eq_epcv}) are
\begin{equation}
\begin{split}
\label{Eq_Vx}
V_{x,c} &= -\frac{1}{A} \sum\limits_{\mathbf{k}'} V^S_{\mathbf{k}-\mathbf{k}'} \rho_{cc}(\mathbf{k}'), \\
V_{x,v} &= -\frac{1}{A} \sum\limits_{\mathbf{k}'} V^S_{\mathbf{k}-\mathbf{k}'} \tilde{\rho}_{vv}(\mathbf{k}').
\end{split}
\end{equation}

Applying a Bogoliubov transformation
\begin{equation}
\begin{pmatrix}
+,\mathbf{k} \\
-,\mathbf{k}
\end{pmatrix}
\equiv
\begin{pmatrix}
\bar{\gamma}_\mathbf{k} \\
\gamma_\mathbf{k}
\end{pmatrix}
=
\begin{pmatrix}
u_\mathbf{k} & -v_\mathbf{k} \\
v^*_\mathbf{k} & u^*_\mathbf{k}
\end{pmatrix}
\begin{pmatrix}
c_{c \mathbf{k}} \\
c_{v \mathbf{k}}
\end{pmatrix},
\end{equation}
which diagonalizes the Hamiltonian into Bogoliubov quasiparticle operators
\begin{equation}
\begin{split}
\hat{\mathcal{H}}_{\rm HF} 
&= \sum\limits_{\mathbf{k}}
\begin{pmatrix}
\bar{\gamma}^\dagger_{\mathbf{k}} & \gamma^\dagger_{\mathbf{k}}
\end{pmatrix} 
\begin{pmatrix}
\varepsilon^{+}_{\mathbf{k}} & 0 \\
0 & \varepsilon^{-}_{\mathbf{k}}
\end{pmatrix}
\begin{pmatrix}
\bar{\gamma}_{\mathbf{k}} \\
\gamma_{\mathbf{k}}
\end{pmatrix}.
\end{split}
\end{equation}
The quasiparticle energies are
\begin{equation}
\begin{split}
\varepsilon^{\pm}_{\mathbf{k}} &= \frac{1}{2}( \varepsilon_{c \mathbf{k}} + \varepsilon_{v \mathbf{k}} )
\pm \sqrt{ \xi^2_\mathbf{k} + |\Delta_\mathbf{k}|^2 }, \\
\end{split}
\end{equation}
where
\begin{gather}
\xi_\mathbf{k} = \frac{1}{2}( \varepsilon_{c \mathbf{k}} - \varepsilon_{v \mathbf{k}} ), \label{Eq_xi}
 \\
\Delta_\mathbf{k} = \frac{1}{A} \sum\limits_{\mathbf{k}'}V^D_{\mathbf{k}-\mathbf{k}'} \rho_{cv}(\mathbf{k}'). \label{Eq_Delta}
\end{gather}

At finite temperature, the density matrix elements take the form
\begin{equation}
\label{Eq_rho}
\begin{split}
\rho_{cc}(\mathbf{k}) &= \langle c^\dagger_{c\mathbf{k}} c_{c\mathbf{k}} \rangle = |v_\mathbf{k}|^2 f(\varepsilon^-_{\mathbf{k}}) + |u_\mathbf{k}|^2 f(\varepsilon^+_{\mathbf{k}}), \\
\tilde{\rho}_{vv}(\mathbf{k}) &= \langle c^\dagger_{v\mathbf{k}} c_{v\mathbf{k}} \rangle - 1 = |u_\mathbf{k}|^2 f(\varepsilon^-_{\mathbf{k}}) + |v_\mathbf{k}|^2 f(\varepsilon^+_{\mathbf{k}}) - 1, \\
\rho_{cv}(\mathbf{k}) &= \langle c^\dagger_{v\mathbf{k}} c_{c\mathbf{k}} \rangle = u^*_\mathbf{k} v_\mathbf{k} \left[ f(\varepsilon^-_{\mathbf{k}}) - f(\varepsilon^+_{\mathbf{k}}) \right],
\end{split}
\end{equation}
with $f(\varepsilon^{\pm}_{\mathbf{k}}) = [e^{(\varepsilon^{\pm}_{\mathbf{k}}-\varepsilon_F)/k_BT} + 1]^{-1} = [1-\tanh((\varepsilon^{\pm}_{\mathbf{k}}-\varepsilon_F)/2k_BT)]/2$ the Fermi-Dirac distribution. Combining Eqs.~(\ref{Eq_VH}, \ref{Eq_Vx}, \ref{Eq_rho}) yields
\begin{equation}
\frac{1}{2} ( \varepsilon_{c \mathbf{k}} + \varepsilon_{v \mathbf{k}} ) = \frac{1}{2}( \varepsilon_{c \mathbf{k}}^{(0)} + \varepsilon_{v \mathbf{k}}^{(0)} ). \\
\end{equation}

\begin{figure*}[!htb]
\centering
\includegraphics[width=1.0\textwidth]{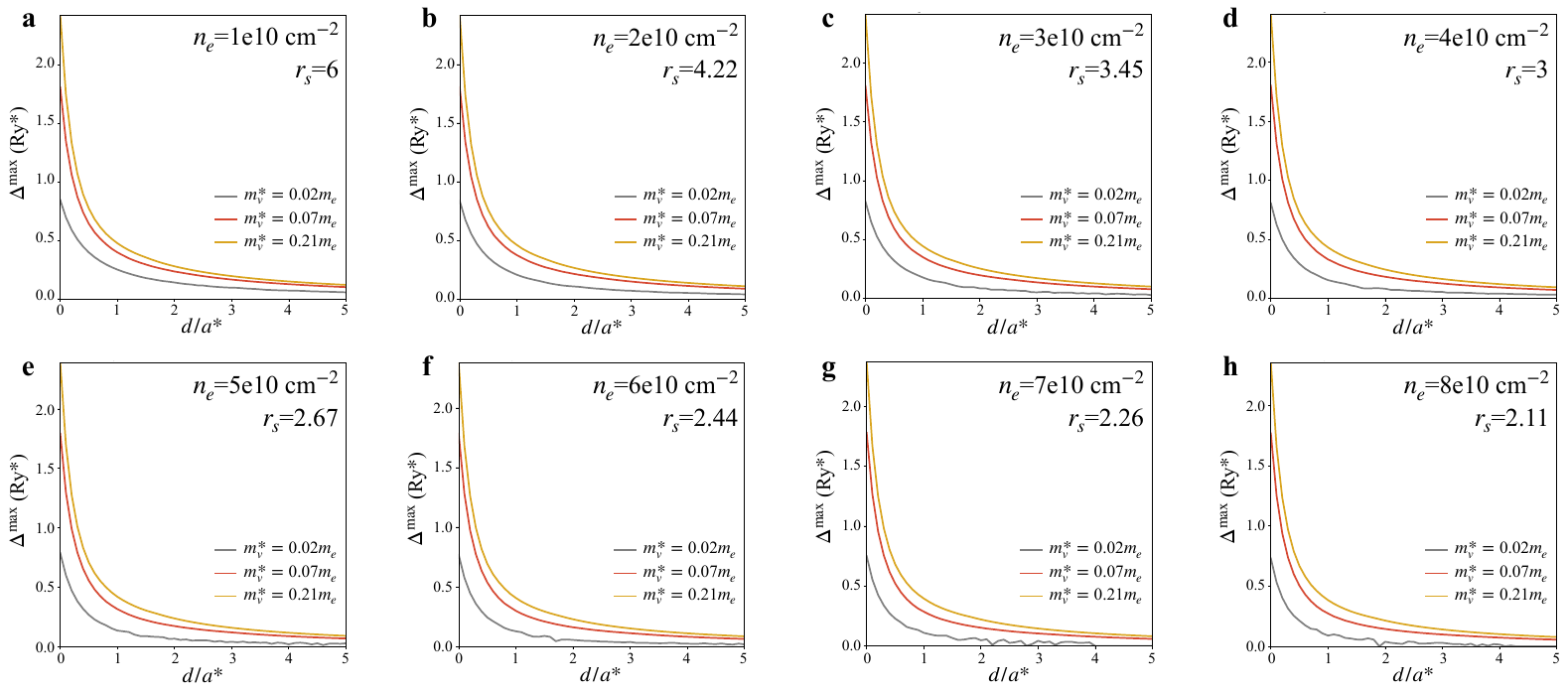}
\caption{\label{fig2_diffmv_line} { {\bf{Self-consistent HF at $T=0$ for unequal and isotropic masses, $m_c^* \neq m_v^*$, with $m_c^* = 0.07 m_e$ fixed.}}
Line cuts from Fig.~\ref{fig1_diffmv_2d} are shown for three representative valence band masses: $m_v^*/m_e = 0.02, 0.07$ and $0.21$. In all cases, the maximum order parameter
$\Delta^{\rm max}$ decreases continuously and exponentially with increasing $d/a^*$.
For small $m_v^*$ and large $d$, $\Delta^{\rm max}$ becomes extremely small, requiring very dense $k$-grid to resolve.
Here, Ry$^*$ and $a^*$ in this figure are defined using $m^* = 0.07 m_e$.
}}
\end{figure*}

The gap equations are solved self-consistently,
\begin{widetext}
\begin{equation}
\label{Eq_eh_sc}
\begin{split}
\Delta_\mathbf{k}
&= \frac{\pi e^2}{A\epsilon} \sum\limits_{\mathbf{k}'} \frac{e^{-d|\mathbf{k}-\mathbf{k}'|}}{|\mathbf{k}-\mathbf{k}'|}
\frac{\Delta_{\mathbf{k}'}}{\sqrt{\xi^2_{\mathbf{k}'} + |\Delta_{\mathbf{k}'}|^2}} \Big(f(\varepsilon^-_{\mathbf{k}'}) - f(\varepsilon^+_{\mathbf{k}'})\Big), \\
\xi_\mathbf{k} &= \frac{1}{2} \big( \varepsilon^{(0)}_{c\mathbf{k}} - \varepsilon^{(0)}_{v\mathbf{k}} \big) + \frac{2\pi e^2 d n_e}{\epsilon} - \frac{\pi e^2}{A\epsilon} \sum\limits_{\mathbf{k}'} \frac{1}{|\mathbf{k}-\mathbf{k}'|} \left[ 1-\frac{\xi_{\mathbf{k}'}}{\sqrt{\xi^2_{\mathbf{k}'}+|\Delta_{\mathbf{k}'}|^2}} 
\Big(f(\varepsilon^-_{\mathbf{k}'}) - f(\varepsilon^+_{\mathbf{k}'})\Big)
\right].
\end{split}
\end{equation}
\end{widetext}
At $T=0$, the density matrix elements in Eq.~(\ref{Eq_rho}) simplify to
\begin{gather}
\rho_{cc}(\mathbf{k}) = \langle c^\dagger_{c\mathbf{k}} c_{c\mathbf{k}} \rangle = |v_\mathbf{k}|^2, \nonumber \\
\tilde{\rho}_{vv}(\mathbf{k}) = \langle c^\dagger_{v\mathbf{k}} c_{v\mathbf{k}} \rangle - 1 = |u_\mathbf{k}|^2  - 1, \\
\rho_{cv}(\mathbf{k}) = \langle c^\dagger_{v\mathbf{k}} c_{c\mathbf{k}} \rangle = u^*_\mathbf{k} v_\mathbf{k}, \nonumber
\end{gather}
leading to the zero-temperature self-consistent equations
\begin{equation}
\begin{split}
\Delta_\mathbf{k}
&= \frac{\pi e^2}{A\epsilon} \sum\limits_{\mathbf{k}'} \frac{e^{-d|\mathbf{k}-\mathbf{k}'|}}{|\mathbf{k}-\mathbf{k}'|}
\frac{\Delta_{\mathbf{k}'}}{\sqrt{\xi^2_{\mathbf{k}'} + |\Delta_{\mathbf{k}'}|^2}}, \\
\xi_\mathbf{k} &= \frac{1}{2} \big( \varepsilon^{(0)}_{c\mathbf{k}} - \varepsilon^{(0)}_{v\mathbf{k}} \big) + \frac{2\pi e^2 d n_e}{\epsilon} \\
& \quad- \frac{\pi e^2}{A\epsilon} \sum\limits_{\mathbf{k}'} \frac{1}{|\mathbf{k}-\mathbf{k}'|} \left( 1-\frac{\xi_{\mathbf{k}'}}{\sqrt{\xi^2_{\mathbf{k}'}+|\Delta_{\mathbf{k}'}|^2}} 
\right). \\
\end{split}
\end{equation}
We have used
\begin{align}
|u_\mathbf{k}|^2 &= \frac{1}{2} \left( 1+ \frac{\xi_\mathbf{k}}{\sqrt{\xi_\mathbf{k}^2 + |\Delta_\mathbf{k}|^2}} \right), \\
|v_\mathbf{k}|^2 &= \frac{1}{2} \left( 1- \frac{\xi_\mathbf{k}}{\sqrt{\xi_\mathbf{k}^2 + |\Delta_\mathbf{k}|^2}} \right), \\
u_\mathbf{k}^*v_\mathbf{k} &= \frac{\Delta_\mathbf{k}}{2\sqrt{\xi_\mathbf{k}^2 + |\Delta_\mathbf{k}|^2}}, \\
f(\varepsilon^-_{\mathbf{k}}) &- f(\varepsilon^+_{\mathbf{k}}) = \tanh \Big( \frac{\sqrt{\xi^2_\mathbf{k}+|\Delta_\mathbf{k}|^2}}{2k_BT} \Big).
\end{align}
We employ the effective Bohr radius $a^*$ and the effective Rydberg Ry$^*$, defined using the effective mass $m^*$
\begin{equation}
a^* = \frac{\epsilon \hbar^2}{m^* e^2}, \quad {\rm Ry}^* = \frac{e^2}{2a^* \epsilon} = \frac{\hbar^2}{2m^* (a^*)^2},
\end{equation}
as fundamental units of length and energy, respectively.

\begin{figure*}
\centering
\includegraphics[width=1.0\textwidth]{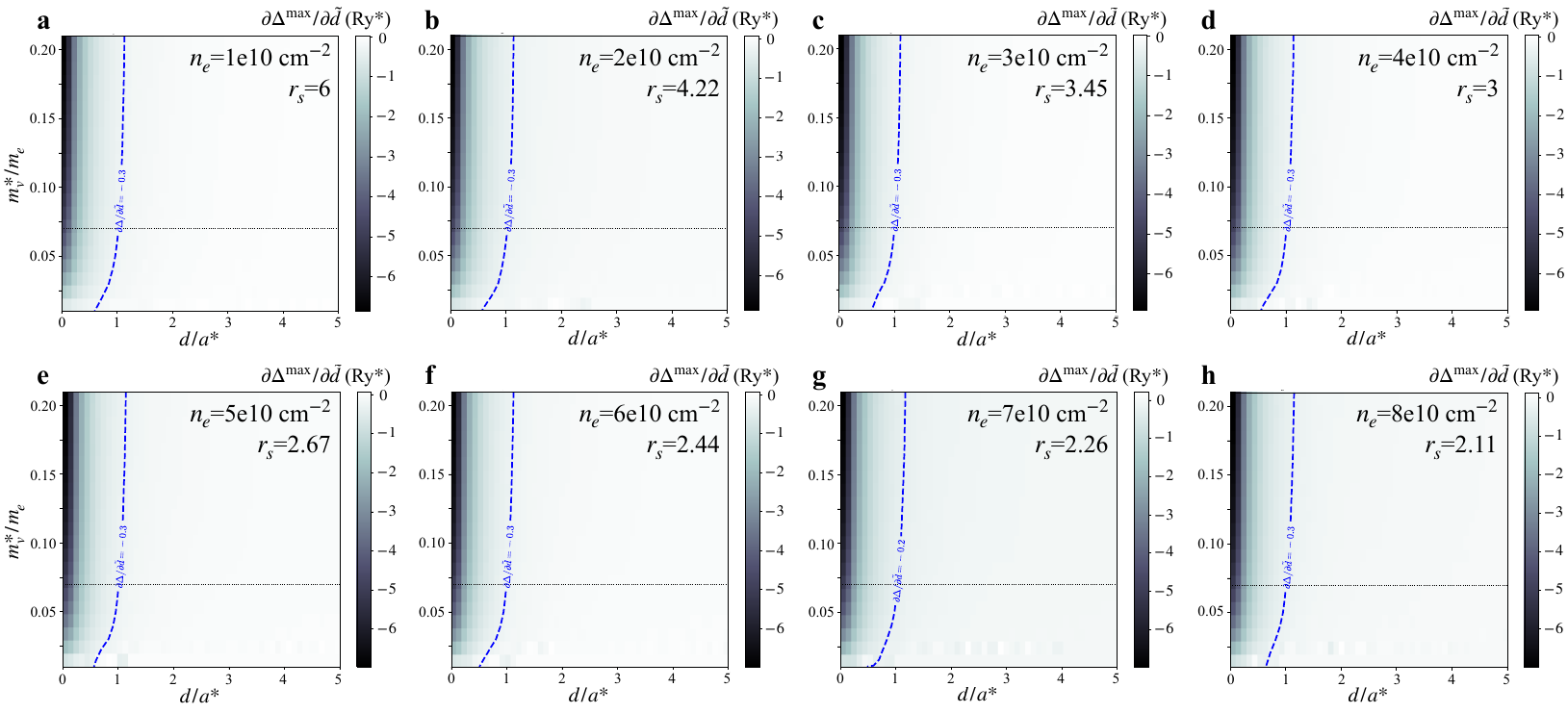}
\caption{\label{fig3_diffmv_dd} { {\bf{$\partial \Delta^{\rm max}/\partial \tilde{d}$ for unequal and isotropic masses, $m_c^* \neq m_v^*$, with $m_c^*=0.07m_e$ fixed.}}
Here, $\Delta^{\rm max}$ is the $T=0$ self-consistent HF order parameter shown in Fig.~\ref{fig1_diffmv_2d}. The blue dashed line in each figure marks the locus where $\partial \Delta^{\rm max}/\partial \tilde{d}$ equals the value for the equal-mass case $m_v^* = m_c^* = 0.07m_e$ at $\tilde{d} = d/a^* = 1$.
}}
\end{figure*}

Figure~\ref{fig1_diffmv_2d} shows the self-consistent HF at $T=0$ for unequal and isotropic masses, $m_c^* \neq m^*_v$, fixing $m^*_c = 0.07 m_e$ and varying $m_v^*$. Over the parameter range in Fig.~\ref{fig1_diffmv_2d},
the maximum order parameter $\Delta^{\rm max} \equiv \max\{ \Delta_{\mathbf{k}}\}$, shown as the colorplot as a function of interlayer distance $d/a^*$ and $m_v^*/m_e$, remains finite, peaking at small $d/a^*$ and exhibiting only weak dependence on $m_v^*$ at fixed $d$.
At higher densities, $\Delta^{\rm max}$ can become exponentially small for small $m_v^*$ and large $d$, requiring extremely fine $k$-grid to resolve.
We select representative cases with $m_v^*/m_e = 0.02, 0.07, 0.21$ (light hole, equal mass, and heavy hole), and plot $\Delta^{\rm max}$ versus $d/a^*$ in Fig.~\ref{fig2_diffmv_line}.
In all cases, $\Delta^{\rm max}$ decreases continuously and exponentially with increasing  $d$, no matter how different $m_c^*$ and $m_v^*$ are.

Mass imbalance does not eliminate superfluidity but suppresses it by lowering its $T_c$. To isolate this effect, we neglect exchange contributions from remote occupied states, setting $m_v^* = \tilde{m}_v^*$ in Eq.~(\ref{Eq_tilde_mv}).
Near $T_c$, the linearized gap equation of Eq.~(\ref{Eq_eh_sc}) is
\begin{equation}
\begin{split}
\Delta_\mathbf{k}
&= \frac{\pi e^2}{A\epsilon} \sum\limits_{\mathbf{k}'} \frac{e^{-d|\mathbf{k}-\mathbf{k}'|}}{|\mathbf{k}-\mathbf{k}'|}
\frac{\Delta_{\mathbf{k}'}}{\xi_{\mathbf{k}'}}\\
&\frac{1}{2}\Big( \tanh\frac{\xi^+_{\mathbf{k}}+|\xi_{\mathbf{k}'}|
-\varepsilon_F}{2k_BT}
- \tanh\frac{\xi^+_{\mathbf{k}}-|\xi_{\mathbf{k}'}|-\varepsilon_F}{2k_BT}\Big), \\
\end{split}
\end{equation}
where $\xi^+_{\mathbf{k}} = ( \varepsilon_{c \mathbf{k}} + \varepsilon_{v \mathbf{k}} )/2 = ( \varepsilon_{c \mathbf{k}}^{(0)} + \varepsilon_{v \mathbf{k}}^{(0)})/2$.

\begin{figure*}
\centering
\includegraphics[width=1.0\textwidth]{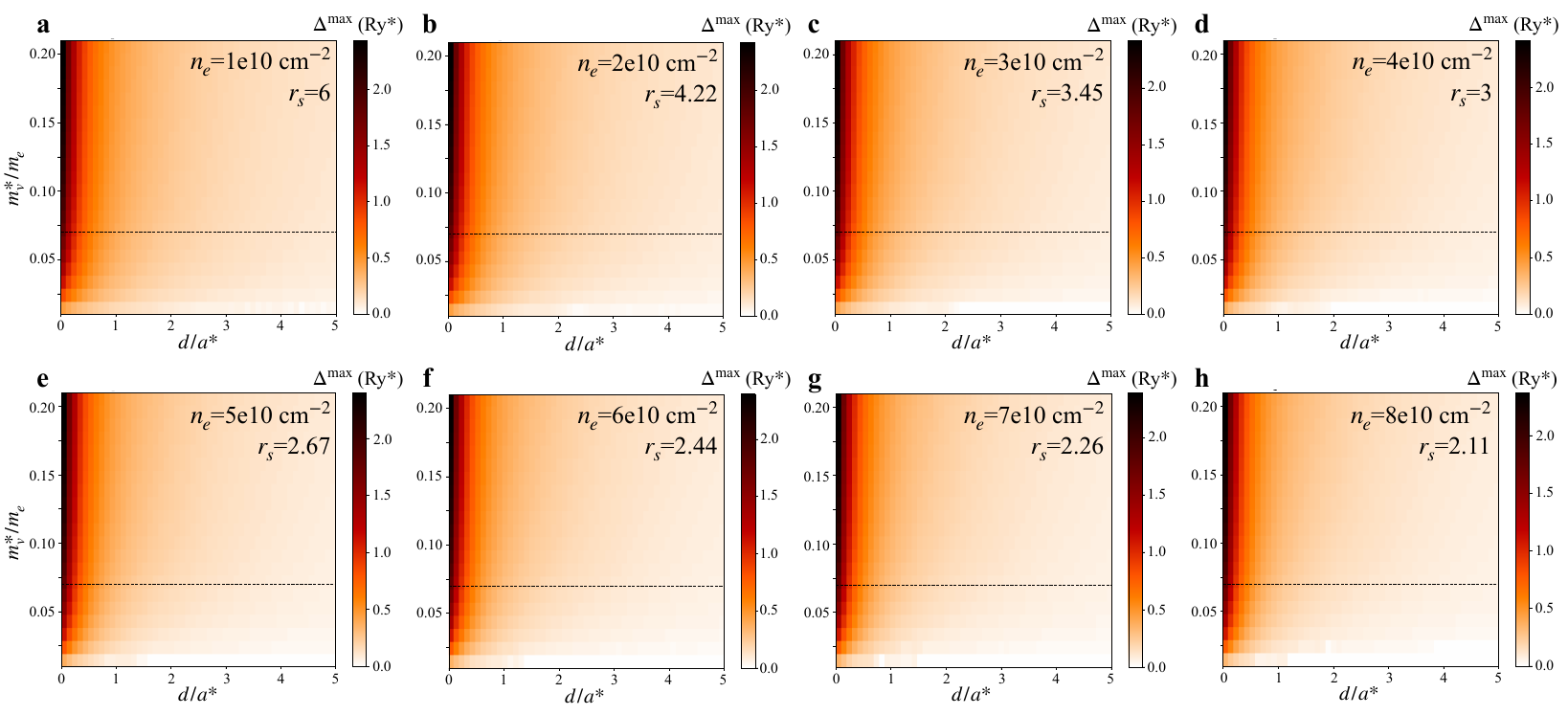}
\caption{\label{fig4_anisom_2d} { {\bf{Self-consistent HF at $T=0$ for average equal but anisotropic masses, $(m_{c,x}^*,m_{c,y}^*) = (m_1^*, m_2^*)$ and $(m_{v,x}^*,m_{v,y}^*) = (m_2^*, m_1^*)$, with $m_2^* = 0.07 m_e$ fixed.}}
Color maps show the maximum order parameter $\Delta^{\rm max} \equiv \max\{ \Delta_{\mathbf{k}}\}$ (in unit of Ry$^*$) as a function of interlayer spacing $d/a^*$ and $m_1^*/m_e$ for electron densities $n_e \in [1, 8] \times 10^{10}$ cm$^{-2}$.
The black dashed line in each figure marks the equal and isotropic mass case $m_1^* = m_2^* = 0.07m_e$.
Here, Ry$^*$ and $a^*$ in this figure are defined using $m^* = 0.07 m_e$.
}}
\end{figure*}

In the equal-mass case, $m_c^*=m^*_v$, $\xi^+ - \varepsilon_F = 0$, leading to
\begin{equation}
\label{Eq_gap_equalm}
\begin{split}
\Delta_\mathbf{k}
&= \frac{\pi e^2}{A\epsilon} \sum\limits_{\mathbf{k}'} \frac{e^{-d|\mathbf{k}-\mathbf{k}'|}}{|\mathbf{k}-\mathbf{k}'|}
\frac{\Delta_{\mathbf{k}'}}{\xi_{\mathbf{k}'}}  \tanh\frac{|\xi_{\mathbf{k}'}|}{2k_BT}. \\
\end{split}
\end{equation}
For unequal masses, $m_c^* \neq m^*_v$, 
\begin{equation}
\tilde{\xi}^+_{\mathbf{k}} \equiv \xi^+_{\mathbf{k}} - \varepsilon_F = \frac{\hbar^2k^2}{4}(\frac{1}{m_c^*} - \frac{1}{m_v^*}) - \frac{E_g}{2} - \varepsilon_F
\end{equation}
Using the identity $\tanh (Y+X) - \tanh(Y-X) = 2\tanh X {\rm sech}^2Y/(1-\tanh^2Y \tanh^2X)$,
\begin{equation}
\begin{split}
&\tanh \frac{\tilde{\xi}^+_{\mathbf{k}} + |\xi_\mathbf{k}|}{2k_BT} - \tanh \frac{\tilde{\xi}^+_{\mathbf{k}} - |\xi_\mathbf{k}|}{2k_BT} \\
&=2\tanh \frac{|\xi_\mathbf{k}|}{2k_BT} \frac{{\rm sech}^2(\frac{\tilde{\xi}^+_{\mathbf{k}}}{2k_BT})}{1-\tanh^2 (\frac{\tilde{\xi}^+_{\mathbf{k}}}{2k_BT}) \tanh^2 (\frac{|\xi_\mathbf{k}|}{2k_BT})},
\end{split}
\end{equation}
the linearized gap equation for unequal masses becomes
\begin{equation}
\label{Eq_Delta_unequalm}
\begin{split}
\Delta_\mathbf{k}
&= \frac{\pi e^2}{A\epsilon} \sum\limits_{\mathbf{k}'} \frac{e^{-d|\mathbf{k}-\mathbf{k}'|}}{|\mathbf{k}-\mathbf{k}'|}
\frac{\Delta_{\mathbf{k}'}}{\xi_{\mathbf{k}'}}  \tanh\frac{|\xi_{\mathbf{k}'}|}{2k_BT} \\
&\qquad \frac{{\rm sech}^2(\frac{\tilde{\xi}^+_{\mathbf{k}}}{2k_BT})}{1-\tanh^2 (\frac{\tilde{\xi}^+_{\mathbf{k}}}{2k_BT}) \tanh^2 (\frac{|\xi_\mathbf{k}|}{2k_BT})}.
\end{split}
\end{equation}
Compared to the equal-mass case Eq.~(\ref{Eq_gap_equalm}), an additional term arises in Eq.~(\ref{Eq_Delta_unequalm}). Since $\tanh^2 (|\xi_\mathbf{k}|/2k_BT) \leq 1$,
\begin{gather}
\frac{{\rm sech}^2(\tilde{\xi}^+/2k_BT)}{1-\tanh^2 (\tilde{\xi}^+/2k_BT) \tanh^2 (|\xi_\mathbf{k}|/2k_BT)} \leq 1.
\end{gather}
Therefore, the kernel in the gap equation of the unequal-mass case is reduced compared to the equal-mass case, leading to a lower $T_c$.
Physically, a nonzero $\tilde{\xi}^+_{\mathbf{k}}$ means that, even with matched Fermi momenta, the pairing partners have different energies away from the Fermi surface, which acts as a pairing-breaking mechanism.
It makes it energetically less favorable to form coherent pairs across the entire relevant momentum space near the Fermi surfaces. This effect is particularly pronounced in the BCS limit where pairing is often delicate and relies on good Fermi surface matching. 
Prior studies of the BCS-BEC crossover in electron-hole bilayers have similarly found that mass imbalance weakens superfluidity \cite{Pieri_BCSBEC_2007, Kaneko_ED_2013, Zittartz_eh_MassAniso_1967, Conti_ehSuperfluid_1998}, especially at higher densities.
Furthermore, $( \varepsilon_{c \mathbf{k}}^{(0)} - \varepsilon_{v \mathbf{k}}^{(0)} )/2 \approx \hbar^2 k^2/{4\mu_r} + E_g/2$, where $\mu_r = m_c^* m_v^*/(m_c^* + m_v^*)$ is the reduced mass, controlling the effective density of states. For a fixed average mass, $\mu_r$ is maximized when $m_c^*=m^*_v$. Therefore, mass asymmetry reduces $\mu_r$, reducing the effective density of states contributing to pairing and suppressing both $T_c$ and $\Delta^{\rm max}$.

Figure~\ref{fig3_diffmv_dd} plots the derivative of $\Delta^{\rm max}$ with respect to $\tilde{d} \equiv d/a^*$, $\partial \Delta^{\rm max}/\partial \tilde{d}$, to track the BEC-BCS crossover, the strong coupling (small $d$) to weak coupling (large $d$) crossover behavior. For equal and isotropic masses, this crossover occurs at $\tilde{d} \sim 1$, consistent with the exciton radius $a^*$ defined by the identical effective mass. For unequal masses, the crossover shifts to a different value of $\tilde{d}$ since the exciton size depends on both effective masses.

\begin{figure*}
\centering
\includegraphics[width=1.0\textwidth]{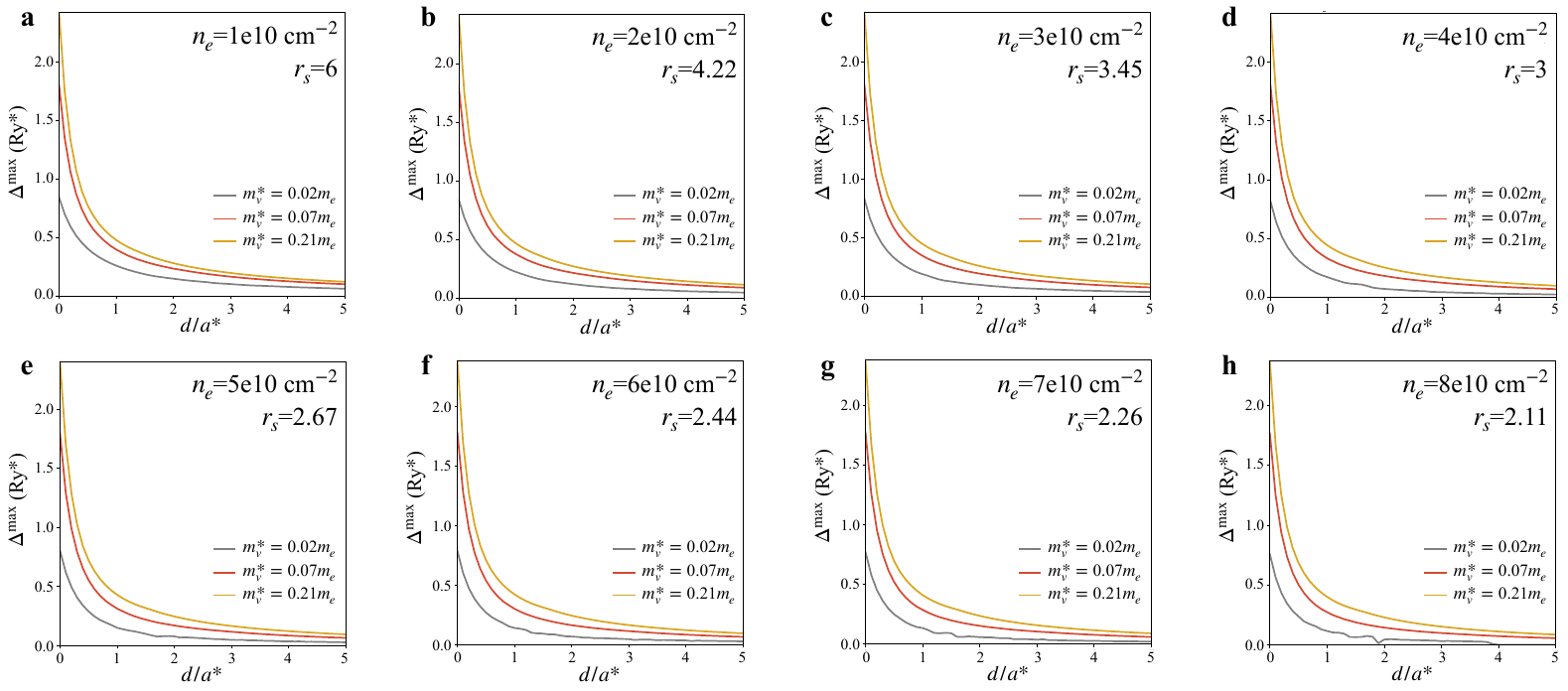}
\caption{\label{fig5_anisom_line} { {
\bf{Self-consistent HF at $T=0$ for average equal but anisotropic masses, $(m_{c,x}^*,m_{c,y}^*) = (m_1^*, m_2^*)$ and $(m_{v,x}^*,m_{v,y}^*) = (m_2^*, m_1^*)$, with $m_2^* = 0.07 m_e$ fixed.}}
Line cuts from Fig.~\ref{fig4_anisom_2d} are shown for three representative $m_1^*$ values, $m_1^*/m_e = 0.02, 0.07$ and $0.21$.
In all cases, the maximum order parameter $\Delta^{\rm max}$ decreases continuously and exponentially with increasing $d/a^*$.
The features are similar to unequal and isotropic masses case in Fig.~\ref{fig2_diffmv_line}.
Here, Ry$^*$ and $a^*$ in this figure are defined using $m^* = 0.07 m_e$.
}}
\end{figure*}

\section{HF with equal but anisotropic masses}
\begin{figure*}
\centering
\includegraphics[width=1.0\textwidth]{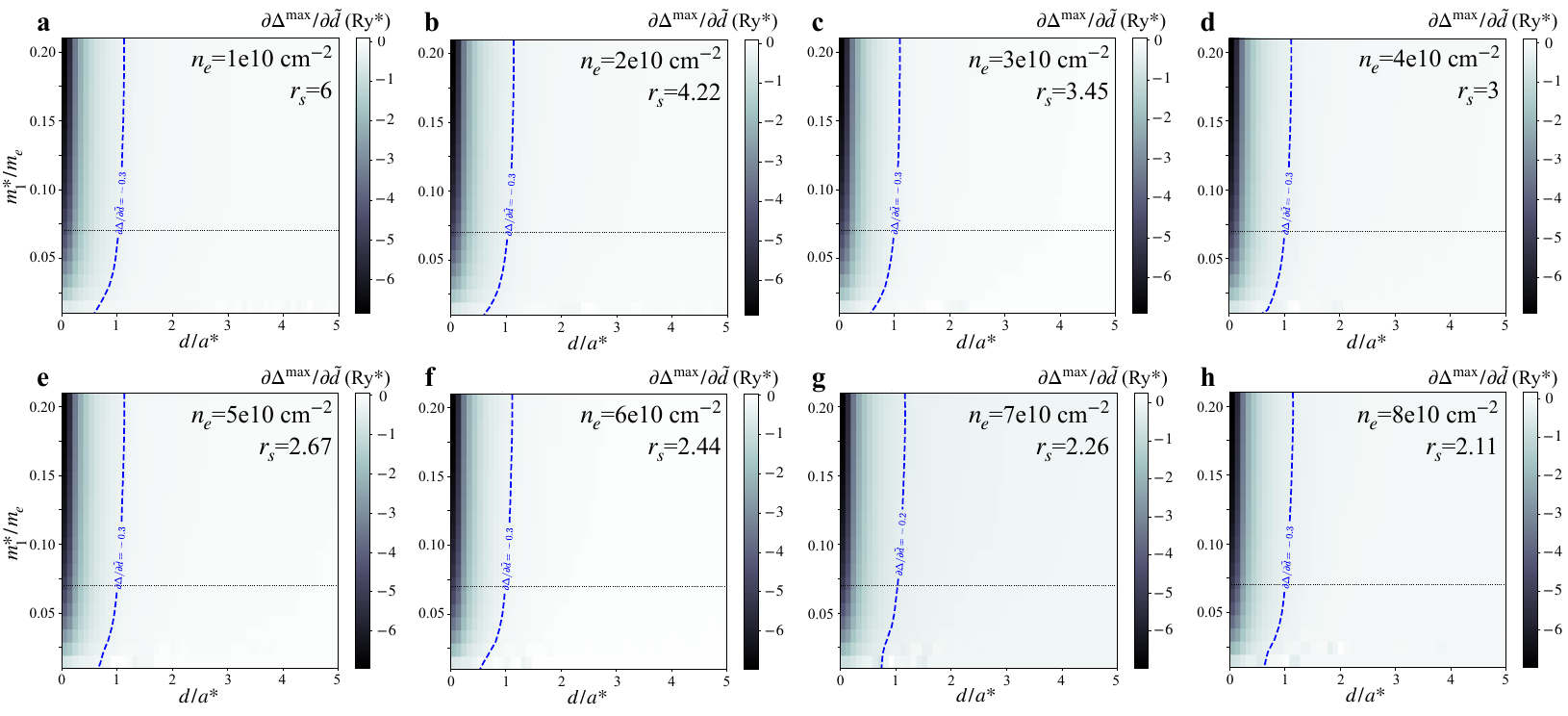}
\caption{\label{fig6_anisom_dd} { {\bf{$\partial \Delta^{\rm max}/\partial \tilde{d}$ for average equal but anisotropic masses, $(m_{c,x}^*,m_{c,y}^*) = (m_1^*, m_2^*)$ and $(m_{v,x}^*,m_{v,y}^*) = (m_2^*, m_1^*)$, with $m_2^* = 0.07 m_e$ fixed.}}
Here, $\Delta^{\rm max}$ is the $T=0$ self-consistent HF order parameter shown in Fig.~\ref{fig4_anisom_2d}. The blue dashed line in each figure marks the locus where $\partial \Delta^{\rm max}/\partial \tilde{d}$ equals the value for the equal-mass case $m_v^* = m_c^* = 0.07m_e$ at $\tilde{d} = d/a^* = 1$.
}}
\end{figure*}

We now examine the case where conduction and valence bands have equal average mass but opposite anisotropies. Specifically, the conduction band has $(m^*_{c,x},m^*_{c,y}) = (m^*_1, m^*_2)$, while the valence band has $(m^*_{v,x},m^*_{v,y}) = (m^*_2, m^*_1)$:
\begin{equation}
\label{Eq_aniso}
\begin{split}
\varepsilon^{(0)}_{c \mathbf{k}} &= \frac{\hbar^2}{2} \left(\frac{k_x^2}{m_1^*} + \frac{k_y^2}{m_2^*} \right), \\
\varepsilon^{(0)}_{v \mathbf{k}} &= -\frac{\hbar^2}{2} \left(\frac{k_x^2}{\tilde{m}_2^*} + \frac{k_y^2}{\tilde{m}_1^*} \right) -E_g.
\end{split}
\end{equation}
The renormalized mass $\tilde{m}^*$ again includes exchange contributions from remote occupied states
\begin{equation}
\begin{split}
-\frac{\hbar^2}{2} \left(\frac{k_x^2}{\tilde{m}_2^*} + \frac{k_y^2}{\tilde{m}_1^*} \right)
&= -\frac{\hbar^2}{2} \left(\frac{k_x^2}{m_2^*} + \frac{k_y^2}{m_1^*} \right) \\
&\quad \ - \frac{1}{A} \sum\limits_{\mathbf{k}'} V^{S}_{\mathbf{k}-\mathbf{k}'} \rho_{vv}^0(\mathbf{k}').
\end{split}
\end{equation}
The spinless HF Hamiltonian and matrix elements remain as in Eqs.~(\ref{Eq_HamilHF}, \ref{Eq_epcv}, \ref{Eq_VH}, \ref{Eq_Vx}).
We solve the gap equations Eq.~(\ref{Eq_eh_sc}) self-consistently with anisotropic band dispersions Eq.~(\ref{Eq_aniso}). The results are presented in Figs.~\ref{fig4_anisom_2d}-\ref{fig6_anisom_dd}.
As in the unequal and isotropic mass case in the previous section, the maximum order parameter $\Delta^{\rm max}$ decreases continuously and exponentially with increasing interlayer spacing $d$, regardless of the degree of mass anisotropy. Mass anisotropy weakens the pairing strength and suppresses the critical temperature by breaking perfect Fermi-surface nesting through geometric distortion of the Fermi surfaces. It also modifies the exciton size, leading to a shift in the BEC-BCS crossover, as shown in Fig.~\ref{fig6_anisom_dd}.


\section{Conclusion}
We have analyzed electron–hole superfluidity in 2D bilayers with unequal and anisotropic effective masses using a zero-temperature, self-consistent HF approach. Two generic scenarios were considered: isotropic but unequal conduction and valence band masses, and equal average masses with orthogonal in-plane anisotropies. Both mass imbalance and anisotropy were found to weaken pairing by breaking Fermi-surface nesting and reducing the density of states, leading to a lower inferred $T_c$, and to alter the exciton size and shifted the BEC–BCS crossover. In both cases, however, the condensate remained stable across the full range of electron densities and interlayer separations studied, with no transition to an unpaired electron-hole plasma in the absence of screening. These results establish a baseline for assessing the combined influence of mass mismatch and anisotropy in realistic bilayer systems, including van der Waals heterostructures and anisotropic 2D semiconductors.

Our work also establishes that the Fermi surface nesting is not necessary for the bilayer excitonic superfluidity, and the system is always a superfluid condensate at $T=0$ except for undergoing a crossover from the strong-coupling small-separation BEC (with higher $T_c$) to weak-coupling large-separation BCS (with lower $T_c$) phase.  Note that the interlayer separation is large ($d \gg a^*$) or small ($d \ll a^*$) in the dimensionless units involving the effective Bohr radius.  The effective $T=0$ quantum phase diagram depends on several independent dimensionless parameters: $d/a*$, $r_s$, as well as the effective mass ratios.  Such multidimensional phase diagrams are challenging to visualize intuitively, leading to our providing many figures to emphasize the results.

We mention that most of the existing direct experimental evidence for bilayer electron-hole superfluidity is in quantum Hall bilayers, where early work was interpreted in terms of a quantum phase transition as a function of layer separation, primarily because the BEC-BCS crossover is very sharp with $T_c$ suddenly decreasing exponentially once the separation is larger than the typical Landau radius (which for the quantum Hall case replaces the Bohr radius $a^*$ in our zero-field theory).
But recent experiments show evidence for a crossover \cite{XLiu_crossover_2022, Eisenstein_precursor_2019} exactly as we find in our work, but more work would be needed to pin down how the transition changes with the interlayer separation.
However, once $T_c$ drops below the experimental electron temperature ($\sim 40$ mK), the BEC-BCS crossover in our theory will appear to be a quantum phase transition. We believe that the early theoretical work concluding the existence of a quantum phase transition at a specific critical layer separation indicates the softening of a collective mode at a specific momentum \cite{Fertig_1989}, and as such implies a transition to a 2D Wigner crystal or charge density wave, implying that the bilayer U(1) symmetry-broken coherent phase becomes a supersolid rather than a superfluid \cite{Eisenstein_precursor_2019, Scarola_SDS_2005}.
Interestingly, there is recent experimental evidence supporting the possible existence of such a predicted blayer superfluid to supersolid (where both U(1) and translation symmetries are spontaneously broken) transition in graphene under quantum Hall conditions in an applied magnetic field \cite{YZeng_supersolid_2023}.

\section{Acknowledgments}
We thank J. D. Sau, J. P. Eisenstein, Philip Kim, B. I. Halperin, H. A. Fertig, A. H. MacDonald, Enrico Rossi
, and Kun Yang for discussions.
This work is supported by the Laboratory for Physical Sciences.


%

\end{document}